\documentstyle[aps,prl,multicol,epsf]{revtex}
\begin{document}

\title { Role of Inter-Electron Interaction in the Pseudo-Gap Opening
in High T $_c$ Tunneling Experiments}

\author{ A.M.Cucolo$^a$, M.Cuoco$^b$}

\address {Unit\`a INFM, Universit\`a di Salerno, Baronissi (SA) 84081 Italy, 
          $^a$ Dipartimento di Fisica,$^b$ Dipartimento di Scienze Fisiche ''E.R.Caianiello''}

\author{ A.A.Varlamov}

\address {Forum, Institute of Solid State Theory of INFM, Dipartimento di Fisica,
          Universit\`a di Firenze, 50125 Firenze, Italy\\and\\Department of
        Theoretical Physics, Moscow Institute for Steel and Alloys, Leninski pr. 4,
        Moscow 117936, Russia}

\maketitle
  

\begin{abstract}
The analysis of tunneling experiments showing the pseudogap type behavior is
carried out based on the idea of the renormalization of density of states
due to the inter-electron interaction in the Cooper channel (superconducting
fluctuations contribution in tunneling current). It is demonstrated that the
observed kink of the zero-bias conductance $G(0,T)$ of $YBaCuO/Pb$ junctions
in the vicinity of $T_c$ can be explained in terms of fluctuation theory in
a quite wide range of temperature above $T_c$, using the values of
microscopic parameters of the $YBaCuO$ electron spectrum taken from
independent experiments. The approach proposed also permits to explain
qualitatively the shape of the tunneling anomalies in $G(V,T)$ and gives a
correct estimate for the pseudogap position and amplitude observed in the
experiments on $BiSrCaCuO$ junctions.
\end{abstract}

\vskip 0.5truecm


\begin{multicols}{2}
\narrowtext

One of the most currently debated problems of the physics of high
temperature superconductors (HTS) is the interpretation of pseudogap type
phenomena which have been observed in a wide range of oxygen concentrations
and temperatures in the normal state of these materials \cite{l96}. In
underdoped compounds such phenomena are well pronounced and have been
investigated by various probes that include photoemission \cite{LOESSER},
NMR \cite{WARREN}, transport \cite{HWANG}, neutron scattering \cite{MIGNOD},
and optical \cite{ROTTER} measurements. In addition to these, recently the
observation of pseudogap structures in the conductance behavior of the $%
BiSrCaCuO$ based tunnel junctions in a wide range of temperature above $T_c$
has been reported. These measurements were obtained by means of traditional
electron tunneling spectroscopy \cite{TFW97} as well as by STM measurements 
\cite{RENNER,MSW97} and interlayer tunneling spectroscopy \cite{SKN97}. The
remarkable experiments of Ref. \cite{RENNER} give also evidence for
pseudogap existence in the ''normal'' state of the overdoped (metallic)
samples what can be found quite surprising from the point of view of an
ordinary Fermi liquid approach.

To summarize these features, a model independent phase diagram has been
proposed \cite{EMERY}, but there is still no consensus on the microscopic
mechanism behind it. One common line of reasoning is that the normal state
has pseudogap anomalies arising from pairing correlations in a state without
phase coherence \cite{RANDERIA}. In the framework of this model, the
pseudogap is a precursor to the superconducting gap as showed in ARPES\ data 
\cite{LOESSER}. A different class of approaches to understand pseudogap
phenomena is related to antiferromagnetic correlation scenarios \cite{PINES}%
, where the pseudogap has no connection with the superconducting energy gap.
Another point of view involves ''spin-charge separation''. Here the
pseudogap is associated with pairing of $S=\frac 12$ charge neutral fermions
called spinons. This scenario has its origin in the resonating valence bond
(RVB) ideas \cite{ANDERSON}, that were further developed \cite{KOTLIAR} to
understand not only the pseudogap phase, but the entire phase diagram from a
Mott insulator to the overdoped regime.

The purpose of this article is to demonstrate that the account of the
interelectron interaction (IEI) with small momentum transfer (Cooper
channel) permits to explain the main characteristic features of the
pseudogap type structures observed in the tunneling conductance of HTS in
the metallic part of the phase diagram (over-, optimally or slightly
under-doped compounds, where the Fermi surface is supposed to be well
developed). The effect of the electron density of states (DOS)
renormalization induced by IEI was firstly studied in \cite{ARW70,CCRV90}.
It was provided that near the critical temperature this correction is
singular in $\varepsilon =\ln {\frac T{T_c}}$, sign changing and manifesting
itself in the very narrow range of energies $\delta E_c\sim \sqrt{T_c(T-T_c)}
$ for clean and $\delta E_d\sim T-T_c$ for dirty systems. Nevertheless it
turns out that such DOS renormalization results in the appearence of the
wide pseudogap type structure in the tunneling conductance \cite{VD83}.
Actually, this effect was firstly observed experimentally in conventional $%
Al-I-Sn$ junction \cite{Khachat}. We show below that this analysis can be
extended to HTS tunnel junctions and permits us to fit the well defined
''kink'' around $T_c$ in the zero-bias conductance $G(V=0,T)$ behaviour of $%
YBaCuO$ based junctions \cite{tun,AMC91} as well as to explain the
appearence of the gap-like structures in the $\frac{dI}{dV}$ characteristics
of intrinsic $BiSrCaCuO$ junctions at temperatures above $T_c$ \cite{SKN97}.

By using the Lawrence-Doniach quasi 2D model for the electron spectrum one
can generalize the expression for the fluctuations contribution to the
differential conductance of a tunnel junction (see \cite{VD83,VBML98}) with
one HTS electrode being in the vicinity of $T_c$: 
\begin{eqnarray}
\frac{\delta G_{fl}(V,\varepsilon )}{G_n} &=&\frac 1{2T}\int_{-\infty
}^\infty \frac{dE}{\cosh ^2\displaystyle{\frac{E+eV}{2T}}}\delta
N_{fl}(E,\varepsilon )=  \label{flcon1} \\
\ &=&Gi_{(2)}\ln \left( \frac 2{\sqrt{\varepsilon }+\sqrt{\varepsilon +r}%
}\right) {\rm Re}\psi ^{\prime \prime }\left( \frac 12-\frac{ieV}{2\pi T}%
\right) ,  \nonumber
\end{eqnarray}
where $\psi (x)$ is the digamma function, $r=4\xi _{\perp }^2(0)/s^2$ is the
Lawrence-Doniach anisotropy parameter($\xi _{\perp }$is the c-axis coherence
length, $s$ is the interlayer distance) which controls the dimensional
crossover from 2D to 3D regime. By mean of $Gi_{(2)},$ the 2D
Ginzburg-Levanyuk parameter \cite{VBML98}, characterizing the strength of
fluctuations, is introduced ( in the most interesting for our consideration
clean case $Gi_{(2)}=\frac{T_c}{14\zeta (3)E_F}$).

One can notice that the sharp decrease ($\sim \frac 1\varepsilon $) of the
density of the electron states generated by IEI in the immediate vicinity of
the Fermi level ($\delta E_c\sim T_c\sqrt{\epsilon }$) surprisingly results
in the much more moderate growth of the tunnel conductance at zero voltage ($%
\sim \ln {\frac 1\varepsilon }$) and in the appearance of some type of the
pseudo-gap structure in the energy scale $eV=\pi T\gg T-T_c,$ as reported in
Fig. 1 for different values of the reduced temperature. This striking
contradiction to the habitual idea of the proportionality between the tunnel
conductance and so-called tunneling density of states is the straightforward
result of the calculation in (\ref{flcon1}) of the real convolution of the
renormalized density of states function $\delta N_{fl}(E)$ with the
Fermi-function derivative as the kernel (which cannot be supposed in this
case as $\delta $-function-like kernel ) side by side with the sum rule: $%
\int_0^\infty \delta N_{fl}(E)dE=0.$ The sense of the last identity can be
easily understood: the account of IEI cannot create new electron states, it
can redistribute the existing only. This condition leads to the exact
cancellation of the main order contribution to the tunnel current (singular
in $1/\varepsilon $), originating from the domain $E{<}T_c%
\sqrt{\varepsilon }$, and to the necessity of more delicate treatment of the
tunnel current general expression what was really done in the purpose to
carry out the logarithmic singularity of expression (\ref{flcon1}).

As inset in Fig. 1, the measurements of the differential resistance of an $%
Al-I-Sn$ tunnel junction at temperatures slightly above the critical
temperature of $Sn$ electrode are presented \cite{Khachat}. It is worth
mentioning that the experimentally measured position of the minima at $%
eV\approx 3T_c(Sn)$ is very close to the theoretical prediction $eV=\pi
T_c(Sn)$.

If pseudogap phenomena in HTS are related to a modification of the normal
state DOS, tunnel spectroscopy still remains one of the most powerful tools
to investigate this puzzling aspect of these materials. However, the
tunneling study of HTS is a difficult task because of many reasons. For
instance, the extremely short coherence length requires monolayer-level
perfection at the surfaces which are subject to long oxygen annealings so
that they in general do not satisfy this strict requirement. Another problem
is the bias dependence of the normal background conductance $G_n(V)$ in a
wide range of voltages (up to hundreds of mV) \cite{AMCB}$,$ which is
necessary to scan for the study of the $I-V$ characteristics of HTS
junctions.

In spite of these difficulties, pseudogap structures in HTS tunneling
characteristics have been observed in different $BiSrCaCuO$ based junctions 
\cite{TFW97}-\cite{SKN97}, while to our knowledge, no data have been
reported for the $YBaCuO$ compounds. This fact seems to confirm the more
relevant role that fluctuations play in 2D systems. On the other hand, by an
experimental point of view, it can be due to the fact that $BiSrCaCuO$
junctions are relatively easier to obtain for the better stability of the
oxygen at the surface of this material. Nevertheless, we dispose of very
accurate data obtained on high quality $YBaCuO/Pb$ junctions in which
elastic tunneling processes occur without any interaction in the barrier 
\cite{AMCSOLO}. The appearance of the typical kink in the $G(0)$ vs $T$
dependence in these as well as in other group experiments \cite{AMC94}-
\cite{TCLW} induced us to apply expression (\ref{flcon1}) to calculate
quantitatively the fluctuation contribution to the $YBaCuO$ DOS and to
demonstrate that a satisfactory fitting can be obtained in a quite wide
temperature range around $T_c.$

In Fig. 2(a,b) the experimental data (dots) refer to two different $%
YBaCuO/Pb $ planar junctions, as reported in Refs. \cite{tun,AMC91}. The
theoretical fittings (full lines) for the normalized fluctuation part of the
tunneling conductance at zero bias, $\delta G_{fl}(0,\varepsilon
)/G_n(0,T=140K)$ by means of expression (\ref{flcon1}) at $V=0,$ are also
reported. The junction's critical temperature and the magnitude of the
Ginzburg-Levanyuk parameter, have been taken as fitting parameters.

To this respect, it is worth noticing that the tunneling spectroscopy probes
regions of the superconducting electrodes to a depth of $(2\div 3)\xi $ in
contrast with the resistive measurements which sense the bulk percolating
length. The two critical temperatures can be quite different. In our case
the resistive critical temperatures measured on both the $YBaCuO$ crystals
were $T_c(\rho =0)=$ $91.6~K$ , while ''junction's $T_c"=88.4$ $K$ and $89K$
were obtained from the fitting procedure. This fact indicates that a
slightly oxygen deficient $YBaCuO$ layer (on a scale of $\xi )$ is probed by
the tunneling measurements in both junctions \cite{nota}. However, the
samples are still in the proper (metallic) region of the phase diagram. The
magnitude of the fluctuation correction $|\psi ^{\prime \prime }(\frac
12)|Gi_{(2)}\simeq T_c/E_F$ is equal to$\ 0.025$ for junction (a) and $0.016$
for junction (b), leading to the value of $E_F\simeq 0.3~eV$ which is in the
lower range of the existing estimates ($0.2\div 1.0~eV$).

The value of Lawrence-Doniach parameter $r=0.08$ was taken from the
crossover between 2D and 3D regimes in the in-plane conductivity
measurements analysis \cite{FPFV97}. It is worth mentioning that another
independent definition of $r$ from the analysis of the non-linear
fluctuation magnetization \cite{Buz96} leads to a very similar value $%
r=0.11 $. However, our fittings turn out to be not too sensitive to the
value of $r$ . We observe that the temperature range in which the expression
(\ref{flcon1} ) satisfactorily reproduces the behaviour of the zero-bias
conductance, extends up to 110 K for junction (a) with $T_c=88.4$ $K$ and up
to 105 K for junction (b) with $T_c=89$ $K.$

Now, we would like to discuss more in details the recent experimental
evidence of pseudogap structures in the conductance curves of intrinsic {\it %
BiSrCaCuO} junctions \cite{SKN97}. As it was already mentioned, the IEI
renormalization of the DOS at the Fermi level leads to the appearance of
similar structures with the characteristic maximum position determined by
the temperature $T$ instead of the superconducting gap value: $eV_m=\pi T$.
For the HTS compounds this means a scale of $20-40~meV$, considerably larger
than in the case of conventional superconductors.

In Fig. 3 the experimental data (dots) refer to thin stacks of $%
Bi_2SrCaCu_2O_8$ intrinsic Josephson junctions. In the experiments \cite
{SKN97}, the pseudo-gap opening in $\frac{dI}{dV}$ characteristics was found
for temperatures up to $180~K$. The curve (a) in the Figure (full line) is
the theoretical fitting for the 90 K data. In view of the strong anisotropy
of the $BiSrCaCuO$ spectrum, $r=0$ has been assumed, while $|\psi ^{\prime
\prime }(\frac 12)|Gi_{(2)}=0.0085$ and $T_c=87~K$ have been extracted from
the expression (\ref{flcon1}). One can observe that the structure amplitude
is well reproduced with an error of less than 5 $\%$ on the maximum
position. We point out that the evaluation of the maxima positions at $90$ $%
K $, $eV_m=\pi T=25meV,$ is consistent with the experimental results on $%
BiSrCaCuO/Pb$ planar junctions \cite{TFW97} as well as with STM\ data \cite
{SKN97}, but is quite lower than the value of pseudogap position observed on
the experiments of Refs. \cite{MSW97,RENNER}.

In this respect, two important comments are necessary that concern the
limits of applicability of the proposed approach. The first one is related
with the magnitude of the effect. It is clear that expression (\ref{flcon1}%
), being a perturbative result, has to be small, so the criterium of the
theory applicability to the HTS phase diagram is $Gi\ln \frac 1{{\varepsilon 
}}\ll 1$. From the values for the junction $T_c$'s found through the fitting
procedure, one can believe in the increase of $Gi$ with the decrease of the
oxygen concentration from the value of the optimal doping, so concluding
that the role of the IEI increases in the underdoped part of the phase
diagram whose properties have to be discussed in the frameworks of some
different theory, not basing on the Fermi-liquid.

It is also important to discuss the temperature range of the pseudo-gap type
phenomena observability following the proposed approach. The expression (\ref
{flcon1}) is obtained in the mean field region $\ln {\frac T{T_c}}\ll 1$,
neglecting the contribution of the short wave-length fluctuations
and reproduces a maximum in conductance at the value of $eV_m=\pi
T_c(1+\varepsilon )$. Nevertheless, the characteristic slow (logarithmic)
dependence on $\varepsilon $ of the fluctuation correction for 2D systems
permits to believe that the result (\ref{flcon1}) can be qualitatively
extended on a wider temperature range. The study of the high temperature
asymptotic behaviour ($\ln {\frac T{T_c}}\gg 1$) for the $e-e$ interaction
in the Cooper channel \cite{ARV83} demonstrates the appearance of the
extremely slow $\ln {\ln {\frac T{T_c}}}$ dependence which matches $\ln {1/}$%
{$\varepsilon $} in the intermediate region and show the importance of the
interaction effects up to high temperatures. In such way one can consider
the reported $T^{*}\sim 200-300~K$ in the strongly underdoped part of the
phase diagram as the temperature where the noticeable concentration of
short-living ($\tau \sim \frac \hbar {k_BT}$) fluctuation Cooper pairs
firstly manifests itself \cite{VBML98}.

To conclude, the idea to relate the tunneling pseudo-gap type phenomena
observed around T$_c$ with the IEI renormalization effects allows us to fit
quantitatively the experimental data with a minimum of microscopic
parameters ($E_F,r$) consistently with the independent measurements. It is
worth stressing that the developed approach allows us to explain in a unique
way the set of pseudo-gap type phenomena, like the increase of c-axis
resistance, the sign-changing c-axis magnetoresistance, the opening of the
pseudo-gap in c-axis optical conductivity and NMR spectra\cite{VBML98}.

We like to thank Prof. M. Suzuki, Prof. T. Watanabe and Prof. Ch. Renner for
making us familiar with their experimental results before publication. We
are grateful to Prof. B. Altshuler, Prof. G. Balestrino, Prof. A. Barone,
Prof. C. Noce and Prof. M. Randeria for valuable discussions. This work was
partially supported by NATO Collaborative Research Grant $\#CRG941187$ and
INTAS grant \# 96-0452. Dott. M. Cuoco whishes to thank partial support by
the EU via the ''Social European Fund''.

\newpage

Fig. 1 Theoretical sketch of the fluctuation induced resistance zero-bias
singularity for different reduced temperatures. Insert: differential
resistance vs voltage measured in an $Al-I-Sn$ junction \cite{Khachat}, at
temperatures just above $T_c(Sn)$.

Figs. 2(a,b). Theoretical fittings (solid lines) of the measured zero bias
conductance behaviour vs temperature (dots) for two different $YBaCuO/$ $Pb$
junctions$,$ as reported in Refs. \cite{tun,AMC91}, respectively. Inserts:
Experiments on a larger temperature scale.

Fig. 3. Theoretical fittings (solid lines) of the G(V,T) structures (dots)
measured in intrinsic $BiSrCaCuO$ tunnel junctions \cite{SKN97} at $T=90~K$.

\end{multicols}


\begin{thebibliography}{99}
\bibitem{nota}  We remark that in this type of junctions the tunneling
barrier is formed by the oxygen deficient HTS top most layers.
\end{thebibliography}
\end{document}